\documentstyle[aps,prl,twocolumn,epsfig]{revtex}
\def\cleardoubleemptypage{\clearpage\if@twoside \ifodd\c@page\else
    \thispagestyle{empty}\hbox{}\newpage
    \if@twocolumn\thispagestyle{empty}\hbox{}\newpage\fi\fi\fi}

\newcommand{\be}{\begin{equation}}%\extraspace}
\newcommand{\ee}{\end{equation}}
\newcommand{\bea}{\begin{eqnarray}}
\newcommand{\eea}{\end{eqnarray}}

\newcommand{\p}{\partial}
\newcommand{\s}{\sigma}

\newcommand{\rF}{{\rm F}}

\newcommand{\up}{\uparrow}

\newcommand{\rd}{\mbox{d}}
\newcommand{\ri}{\mbox{i}}

\newcommand{\rc}{{\rm c}}
\newcommand{\rs}{{\rm s}}

\begin{document}

\title{Scaling Exponents in the Incommensurate Phase of the
Sine-Gordon  and U(1) Thirring Models}
\author{Emiliano Papa and  Alexei M. Tsvelik \\
{\em{Department of Theoretical Physics, University of Oxford,
1 Keble Road, Oxford, OX1 3NP}}}
\maketitle
\vspace{8mm}

%\address{\rm (Received: )}

\begin{abstract}
%\vspace{00mm}
\par

In this paper we study the critical exponents of the quantum 
sine-Gordon and U(1) Thirring models 
in the incommensurate phase. This  phase appears  when the chemical 
potential $h$ exceeds a critical value and is characterized by a 
 finite density of solitons. The  low-energy sector of this phase is critical 
and is  described by the Gaussian model (Tomonaga-Luttinger liquid) 
with the compactification radius dependent on the soliton density 
and  the sine-Gordon model coupling constant $\beta$.

For a fixed value of $\beta$, we find that the Luttinger parameter 
$K$ is equal to  1/2
at the commensurate-incommensurate transition point
 and approaches the asymptotic value 
$\beta^2/8\pi$ away from it. We describe a possible phase diagram of
the model consisting of an array of  
 weakly coupled chains. The possible phases are Fermi liquid, Spin
Density Wave, Spin-Peierls and Wigner crystal.

\vspace{2mm}
\end{abstract}

\vspace{8mm}
%\section{Introduction}

In this paper we study the critical exponents of the quantum
sine-Gordon (SG) and the U(1) Thirring  models in the presence of soliton 
condensate. The Lagrangian of SG  model is given by
\bea
\label{sg}
L = \int \rd x \left\{\frac{1}{2} \left[(\p_t\Phi)^2 - (\p_x\Phi)^2 \right] 
\right. & - &  h\beta/\pi\p_x\Phi +
\\[2mm]
 &+&\left. \frac{\mu^2}{\beta^2}(\cos\beta\Phi - 1)\right\} \ , 
\nonumber
\eea
where $\mu$ and $\beta$ are real parameters, mass-like and coupling
constant respectively and $h$ is chemical potential.
SG model is closely related to   the U(1) Thirring
model (it describes its (iso)spin sector). The Lagrangian density for the
latter model is given by
\bea
{\cal L} = \ri \bar{\Psi}_{\alpha} \gamma^{\mu} \partial_{\mu} \Psi_{\alpha}
& - &  \frac{1}{4} \sum_{a=1,2,3} g_a j^a _{\mu} j^{a \mu} \quad ,
\\ [3mm] \nonumber 
 g_1 = g_2  & = & 
g_{\perp}\, ,\quad    g_3 = g_{\parallel} \quad ,
\label{U(1)Thirring}
\eea
where $j^a _{\mu} = (1/2) \bar{\Psi}_{\alpha} \gamma^{\mu} \tau^a _{\alpha 
\beta} 
\Psi_{\beta}$ are the (iso)spin currents. The relationship between the
two models is established by bosonization (this procedure and the
relationship between the operators is discussed at length in
\cite{blue}).
The coupling constants $g_{\parallel}$  and $g_\perp$ are related to
$\beta$ and to $\mu$. 

The Hamiltonian of the U(1) Thirring model, which as a model of two species
of fermions possesses central charge $c=2$, can be written equivalently
as a sum of two charge- and spin- Hamiltonians, each having central charge
$c=1$, i.e.
$
H_\rc  =  v_\rc/2  % \frac{v_\rc}{2}
\int \rd x \left(:J J: + :\bar{J} \bar{J}:\right) \, 
$
and 
\bea
\label{spin-sector}
H_\rs  =   \frac{v_\rs}{2}  \int  \rd x \left[ :{\bf J} {\bf J}: \right. & + &
:\bar{{\bf J}} \bar{{\bf J}}:  + \,  g_\parallel :J_z J_z :+ 
\\[1mm] \nonumber
 &+& \left.  g_\perp (:J_x J_x: +:J_y J_y^{}:) \right] \quad .
\eea
In the above equations, $J$, $\bar{J}$ and ${\bf J}$, ${\bf {\bar J}}$ are 
chiral charge- and spin-currents given as usual by $J=R^\dagger_\alpha 
R_\alpha$ and ${\bf J}= R^\dagger_\alpha ({\vec \sigma}_{\alpha \beta}/2) 
R_\beta$ for the right currents and similarly for the left ones.

The U(1) Thirring model may be used as a continuous model of
interacting  lattice
fermions. Here one can imagine two situations. First,  one may use 
the (iso)spin part of the Thirring model  as a model describing 
the charge sector of the Mott insulator. In the  latter case the
currents belong to the charge sector and  the spin sector is gapless. It is natural to impose an additional
requirement of the SU(2) symmetry in the spin sector, which fixes the
corresponding compactification radius $\beta_s = \sqrt{8\pi}$. 
In the different
interpretation the model describes the  continuous limit of lattice
fermions with a gapless charge sector and an anisotropic interaction in
the spin sector. In this case one does not expect any restrictions on
the compactification radii. 

In both cases one has the following relationship between the order
parameter fields and the bosonic fields:
\bea
\label{staggered}
& &\rho_{\mbox{st}}(x)\sim a_\rc \sin \left[\frac{\beta_{\rc} \Phi_{\rc}(x)}{2} 
+ 2 k_{\rF} x \right]
~ \cos \frac{\beta_{\rs} \Phi_{\rs} (x)}{2}  +
\\ [1mm] \nonumber
& & \hspace{38mm}+ \, a_{\rm u} \cos \left[\beta_\rc \Phi_\rc(x)\frac{}{} + 4k_\rF x 
\right] \quad ,
%\\ [1mm] \nonumber
% & & [d_\rho=\beta_\rc^2/16\pi+\beta_\rs^2/16\pi= \underline{K_\rc/2+1/2}]
% \longleftarrow
\\ [3mm]
& &n^z (x) \sim
\cos \left[\frac{\beta_{\rc} \Phi_{\rc} (x)}{2} + 2 k_{\rF} x \right]
~ \sin \frac{\beta_{\rs} \Phi_{\rs}  (x)}{2}  \label{bos-n_z}  \quad ,
%\\ \nonumber
% & & [d_{n^z}=\beta_\rc^2/16\pi+\beta_\rs^2/16\pi=\underline{K_\rc/2+1/2}] 
%\longleftarrow
\\ [3mm]
& &n^{\pm} \sim 
\cos \left[\frac{\beta_{\rc} \Phi_{\rc} (x)}{2} + 2 k_{\rF} x \right]
~\exp \left[ \pm \mbox{i} \tilde{\beta}_{\rs} \Theta_{\rs} (x) / 2\right]
%\\ \nonumber
% & & [d_{n^\pm}=\beta_\rc^2/16\pi+\beta_\rs^2/16\pi=\underline{K_\rc/2+1/2}]
%\longleftarrow
\label{bos-n_xy} \quad ,
\\ [3mm]
&&\Delta_{0} (x) \sim \sin \frac{\tilde{\beta}_{\rc} \Theta_{\rc}
(x)}{2}~ \cos
\frac{\beta_{\rs} \Phi_{\rs} (x)}{2} \quad ,
%\\ \nonumber
% & & [d_{\Delta_{0}}={\tilde \beta}_\rc^2/16\pi+\beta_\rs^2/16\pi=
%\underline{1/2K_\rc+1/2}]
%\longleftarrow
\\[3mm] 
&&{\bf \Delta} (x) \sim \sin \frac{\tilde{\beta}_{\rc} \Theta_{\rc}
(x)}{2}~{ \bf n} (x)  \quad ,
\label{triplet}
%\\ \nonumber
% & &
%[d_{\Delta}=\beta_\rc^2/16\pi+\beta_\rs^2/16\pi=\underline{1/2K_\rc+1/2}]
%\longleftarrow
%\quad ,
\eea
with ${\tilde \beta_\rc} \beta_\rc=8\pi$ and ${\tilde \beta_\rs} 
\beta_\rs=8\pi$.

To be definite, let us stick to the Mott insulator version. In this
case the field $\Phi$ of SG model (\ref{sg}) is $\Phi_\rc$ and 
the $\Phi_\rs$ sector is
gapless with $\beta_s = \sqrt{8\pi}$ and $\beta_{\rc} \equiv \beta$. 
When  $\beta^2 < 8\pi$ the cosine term in Eq.(1) is relevant and 
generates a spectral 
gap in the low-energy sector. However, when the chemical potential $h$
exceeds a certain  
threshold, the system becomes gapless again \cite{Bak}. 

In this paper we discuss  the case $\beta \rightarrow 0$. This case
is seldom discussed in the literature because such limit cannot be
achieved by short-range interactions (recall that for the repulsive Hubbard
model $\beta$ varies between $\sqrt{8\pi}$ (weak repulsion) and 
$\sqrt{4\pi}$ (infinite repulsion). In 
\cite{Frahm} is found that the Luttinger liquid parameter of the charge sector 
$K_\rc$ takes values monotonically increasing
in the interval $1/2<K_\rc=\zeta^2(B)/2<1$  as $u=U/t$ decreases  
(the asymptotic dependence of the dressed charge $\zeta(B)$ on $u$ is also 
given in
\cite{Frahm}). 
%\begin{equation}
%\left\{ \displaystyle \begin{array}{ll}\displaystyle
%\zeta(B) = 1 + \frac{\sin k_0}{\pi u}\ln2
%\quad ,
%\quad\quad\quad\mbox{($u \rightarrow +\infty$)}\ ;
%\displaystyle\quad  \\[2mm]
%%\\
%\displaystyle
%\zeta(B) =\sqrt{2}\left( 1 - \frac{u}{2 \pi \sin k_0}\right)
%\quad,\quad\mbox{($u \rightarrow +0$)} \ ,
%%\quad .
%\end{array}
%\right.
%\label{K-thet-h}
%\end{equation}
%where $k_0=\pi n_\rc$ and $n_\rc$ is the filling.

One can achieve small 
$\beta$ considering long-range interactions (such as unscreened Coulomb
interaction which can be realized, for instance,  in quantum wires). 

The limit of small $\beta$ is very curious from the mathematical
point of view because solitons in SG model  become in this case 
infinitely heavy. Thus  
in the absence of solitons (that is in the insulating phase) 
SG  model becomes a model of a
free massive bosonic field with mass $\mu$. This field can be thought
about as an optical phonon. One may wonder what happens when 
one  forces solitons to appear applying the chemical potential. 
As we shall demonstrate,  
the scaling dimensions in the regime of small $\beta$ are
determined not by the soliton mass, but by the mass of the optical
phonon $\mu$.

Let us describe our results. In the Mott insulator interpretation
where the field $h$ truly corresponds to chemical potential we get
the following. The gapless (incommensurate or doped) phase of SG
model has scaling dimensions dependent on $\beta$ and soliton 
concentration $n_{\rm sol}$ (which is related to deviation of the electron
concentration from half-filling). Looking at scaling dimensions of
various operators, one can identify different regions of the phase
diagram. 

The most singular operators are components of 
the staggered magnetization $n^z, n^{\pm}$, the $2k_\rF$ and the $4k_\rF$
components of the staggered charge density  $\rho_{\mbox{st}}$.
The first four operators  all have the same scaling dimension
\be
d_n = \frac{1}{2} + \frac{K_\rc}{2}
\quad ,
\ee
where $K_\rc(n_{\rm sol},\beta)$ varies from 1/2 at $n_{\rm sol} \rightarrow 0$ 
to $\beta^2/8\pi$ in the limit of large chemical potential. The $4k_F$ charge 
density component has scaling dimension 
\be
d_{4k_\rF} = 2K_\rc
\quad .
\ee
Therefore $d_n,d_{4k_\rF}  < 1$ in the entire
area of interest and the corresponding
susceptibilities remain singular throughout this area. However,
starting from $K_\rc < 1/3$ (with $d_{4k_\rF}< d_n$) the $4k_F$-component is 
more singular. Its
correlation function decays more slowly being also oscillatory with
wavelength $\lambda=2\pi/4k_\rF=a$, the intersite spacing. In this case
in the presence of other chains the material will undergo a phase transition to
a Wigner crystal ($4k_F$-Charge Density Wave). For $K_\rc > 1/3$ the material will undergo a phase transition 
either of  Spin Density Wave (SDW) or Spin Peierls (SP) type.
%1/3$) the Wigner crystal ($4k_F$ Charge Density Wave) type.
 The importance
of the single-electron interchain tunneling is determined by  the magnitude 
of the scaling dimension of the  single fermion creation
 (annihilation) operator: 
\be
d_{\psi} = \frac{1}{8}\left(\sqrt {K_\rc} + 1/\sqrt {K_\rc}\right)^2
\quad .
\ee
Since the operator describing single-electron interchain tunneling has
scaling dimension $2d_{\psi}$, it becomes relevant at 
$K_\rc > \sqrt 2 - 1$.

From the exact solution \cite{JNW,zam} the excitation spectrum is given 
by
\begin{eqnarray}
E = \epsilon(\theta) \quad , \quad  P = 2\pi\int_0^{\theta}
\rd \theta'\sigma(\theta')
\quad ,
\end{eqnarray}
where $\epsilon(\theta)$ is obtained from the solution of the following
Bethe-ansatz integral equation:
\bea
\epsilon(\theta) + \int_{-B}^B \rd \theta' G(\theta -
\theta')\, \epsilon(\theta') = M_{\rm s}\cosh\theta - h 
\quad .
\label{integ}
\eea
$M_\rs$ is the kink's mass and the Fourier image of the kernel is
\be
G(\omega) = \frac{
\sinh\frac{\pi \omega(1-2\tau)}{2(1-\tau)}}
{2\cosh(\frac{\pi\omega}{2})\sinh\frac{\pi\omega \tau}{2(1-\tau)}} 
\quad, \quad  \tau = \frac{\beta^2}{8\pi} \quad.
\label{kernel}
\ee
The soliton's mass is a function of the coupling constant $\beta$ and at small 
values of $\beta$ it grows like $1/\tau$, where $\tau$ here and in the 
following is the short notation for $\beta^2/8 \pi$. A plot of  $M_\rs$ versus  $\tau$ can be found in \cite{PT1}.

The value of  $B$  in Eq.~(\ref{integ}) is determined by the condition
$\epsilon(\pm B)=0$,
and depends on the coupling constant $\beta$  and  the field $h$.

\begin{figure}
\unitlength=1mm
\begin{picture}(80,53)
%\put(-1.5,3){\epsfig{file=IQzt1.MET,height=50mm}}
%\put(-1.5,3){\epsfig{file=IQet4.MET,height=50mm}}
\put(-1.5,3){\epsfig{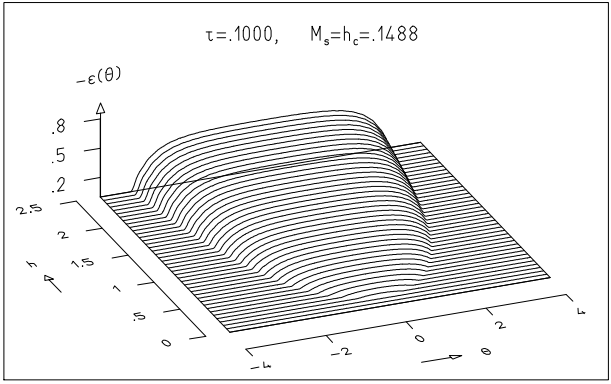}}
\end{picture}
\caption{Plot of $-\epsilon(\theta)$ on $\theta$ and $h$ dependence
obtained from the
numerical solution of Eq.~(\ref{integ}),
only inside the interval ($-B,B$), is shown. 
The coupling constant $\tau$ is taken to have value 0.1.
% but one can think of
%it as having kept $h$ fixed and increased $\beta^2 \sim \tau$.}
}
\label{fig1}
\end{figure}

At the point $\tau = 1/2$ $(\beta^2 = 4\pi)$ the kernel (\ref{kernel}) 
vanishes 
and $\epsilon(\theta)$ is given simply by the right-hand side of (\ref{integ}).
For other values of $\tau$,   Eq.~(\ref{integ}) can be solved numerically.
A plot of $\epsilon(\theta)$, for a fixed value of the coupling constant
$\beta$ and different values of the field $h$, is shown on Fig.~\ref{fig1}.
%can be found in \cite{PT1}.

The soliton condensate  appears when the  field $h$ exceeds the critical value,
found from (\ref{integ}) to be $h_\rc=M_\rs$.  
The soliton contribution to the 
ground state energy of the sine-Gordon model is given by
\be
{\cal E}_0 = \frac{M_\rs}{2 \pi} \int_{-B}^B {\rm d} \theta
\cosh\theta \,  \epsilon(\theta)
\quad .
\label{ground-st}
\ee
 The average number of
solitons  $n$  and  the susceptibility $\chi$ are given by 
%the number of particles in the ground state (in the upper band) will be
\begin{equation}
n_{\rm sol} = - \frac{\partial {\cal E}_{\rm 0}}{\partial h}  \quad , \quad 
\chi = \frac{\partial n_{\rm sol}}{\partial h} 
\quad .
\label{chih}
%= - \frac{2\tau}{\pi}\frac{\partial^2 \tilde{\cal F}}{\partial Q^2}
%= - \frac{1}{2\pi}\frac{\partial \tilde{\cal F}}{\partial Q}
\end{equation}

One can also calculate the number of solitons using the density 
distribution function $\s(\theta)$:
\bea
n_{\rm sol} & = & \int_{-B}^B \rd \theta \, \sigma(\theta) \quad , 
\eea
with $\sigma(\theta)$  satisfying the following integral equation:
\bea
\sigma(\theta) & + & \int_{-B}^{B} \rd \theta' G(\theta - \theta') \, 
\sigma(\theta') = M_{\rs}\cosh\theta/2\pi
\quad .
\label{integ2}
\eea

% The excitation spectrum is given by
%\begin{eqnarray}
%E = \epsilon(\theta), ~~ P = 2\pi\int_0^{\theta}d\theta'\sigma(\theta')
%\end{eqnarray}

At low-energies the  excitation spectrum can be linearized at the Fermi points 
$\theta = \pm B$, with the Fermi velocity given by:
\begin{eqnarray}
v_\rF = \left. \frac{\partial\epsilon(\theta)}{\partial\theta}\right|_{\theta =
B}\frac{1}{2\pi\sigma(B)}
\quad .
\end{eqnarray}
The boundary $B$ is the same as the one of Eq.~(\ref{integ}).

%The susceptibility:
%\begin{equation}
%\chi = \frac{\partial n}{\partial h} = -
%\frac{2\tau}{\pi}\frac{\partial^2 \tilde{\cal F}}{\partial Q^2}
%\end{equation}

The scaling exponents of the Luttinger liquid are expressed in terms of a
single parameter $K$. This parameter can be calculated in two different
ways. 

First of all, it is related to the dressed charge $\zeta(B)$ which can be
found from the solution of  the following integral equation  \cite{Frahm}:
%\cite{kawakami}:
\begin{eqnarray}
\zeta(\theta) + \int_{-B}^B \rd \theta' G(\theta - \theta')\, \zeta(\theta') = 1
\quad ,
\label{integ0}
\end{eqnarray}
where $G(\theta)$ and $B$ are the same as in Eq.~(\ref{integ}).
The parameter $K$ is related to $\zeta(B)$  by 
\begin{equation}
K = c [\zeta(B)]^2
%\quad ,
\label{integ00}
\end{equation}
up to a constant factor which, as results from (\ref{K-asympt}) and 
(\ref{K-asympt2}), is $c=1/2$. The scaling dimensions are
\begin{equation}
\Delta_{n,m} = \frac{1}{16}\left[n \zeta(B) \pm \frac{2 m}{\zeta(B)}\right]^2
%\Delta_{n,m} = \frac{1}{16}[n \zeta(B) \pm 2 m/\zeta(B)]^2
\quad .
\end{equation}

Another way to determine $K$ is to use the identity relating $K$ to the
susceptibility $\chi$ and the Fermi velocity $v_\rF$ \cite{Frahm,kawakami}:
\be
K = \frac{1}{2}\pi\chi v_\rF
\quad .
\label{K-chi}
\ee
We can use the above formula as a check of the result obtained by
(\ref{integ00}). 

First, consider the case $h \rightarrow + \infty$ where the limit $B$ goes
to infinity. In this limit the integral equation Eq.~(\ref{integ}), as
$\theta\rightarrow B$, can be written in the following form 
\bea
\epsilon(\theta) + \int_{-B}^B \rd \theta' G(\theta -
\theta')\, \epsilon(\theta') = M_{\rm s}\frac{e^\theta}{2} - h
\quad ,
\eea
and the derivative $\epsilon'(\theta)={\rm d} \epsilon(\theta)/{\rm d} \theta$,
taking into account the boundary conditions for $\epsilon(\theta)$, fulfills
\bea
\epsilon'(\theta) + \int_{-B}^B \rd \theta' G(\theta -
\theta')\, \epsilon'(\theta') = M_{\rm s}\frac{e^\theta}{2} 
\quad .
\label{integ3}
\eea
As can be seen from Eq~(\ref{integ2}) and Eq.~(\ref{integ3}), 
$\epsilon'(\theta)$ and $2\pi \sigma(\theta)$ fulfill the same integral 
equation, as $\theta\rightarrow +\infty$, meaning that the 
Fermi velocity in the limit $h\rightarrow +\infty$ goes to 1, 
$v_\rF \rightarrow 1$.

According to \cite{zam} the ground state energy has the
following $h$-asymptotic behavior:
\begin{equation}
{\cal E}_0 = - \tau h^2/\pi, ~~ n_{\rm sol} = 2 \tau h/\pi, ~~\chi = 
\beta^2/(2\pi)^2,
%\tilde{\cal F} = - Q^2/2, ~~ n = Q/2\pi, ~~\chi = \beta^2/(2\pi)^2,
\end{equation}
which, for the Luttinger liquid parameter, gives 
\begin{equation}
K = \beta^2/8\pi =\tau 
\quad .
\label{K-asympt}
\end{equation}
On the other hand, we can derive the same result from Eq.~(\ref{integ0}).
In the limit $B \rightarrow + \infty$ this equation can be written
as a Wiener-Hopf equation:
\begin{equation}
\zeta(\theta) + \int_{-\infty}^0 \rd \theta' G(\theta - \theta') \, 
\zeta(\theta') = 1 \label{zeta}
\end{equation}
(We have shifted $\theta$ by $B$, so that now we need to calculate
$\zeta(0)$).
The solution for
\[
\zeta^{(-)}(\omega) = \int_{-\infty}^0 \rd \theta \,
{\rm e}^{\ri \theta\omega} \, \zeta(\theta)
\]
is
\begin{equation}
\zeta^{(-)}(\omega) = \frac{1}{G^{(+)}(0)G^{(-)}(\omega)(\ri \omega + 0)}
\label{hpf}
\end{equation}
where
\[
1 + G(\omega) = G^{(+)}(\omega)G^{(-)}(\omega)
\]
with $G^{(\pm)}(\omega)$ being analytic in the upper (lower) half-plane and
$G(\infty) = 1$.
The limit at $\theta = 0$ is given by
\begin{figure}
\unitlength=1mm
\begin{picture}(80,53)
%\put(-1.5,3){\epsfig{file=IQzt1.MET,height=50mm}}
\put(-1.5,3){\epsfig{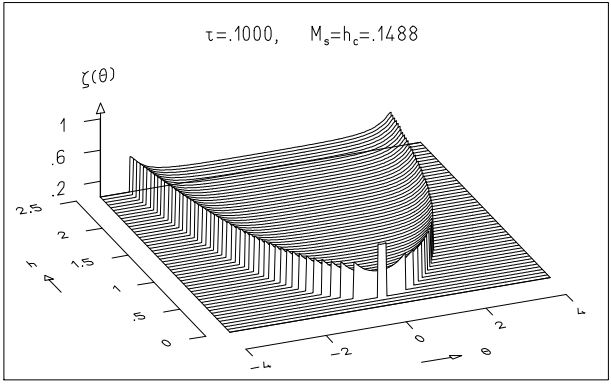}}
\end{picture}
\caption{Plot on $\theta$ and $h$ dependence of the
dressed charge $\zeta(\theta)$ of Eq.~(\ref{integ0}), only inside the
interval
($-B,B$), is shown.
The value of the coupling constant $\tau$ is kept fixed.
% but one can think of
%it as having kept $h$ fixed and increased $\beta^2 \sim \tau$.}
}
\label{dresscharge}
\end{figure}
\bea
\nonumber
\zeta(0) & = & \lim_{\omega \rightarrow \infty} \ri \omega\zeta^{(-)}(\omega)
= [G^{(+)}(0)]^{-1} 
\\[3mm]
& = & [1 + G(\omega = 0)]^{-1/2} = \sqrt{2\tau}
\quad .
\eea
Thus we get
\be
K = \frac{1}{2}[\zeta(0)]^2 = \tau = \frac{\beta^2}{8\pi}
\quad ,
\label{K-asympt2}
\ee
which is consistent with Eq.~(\ref{K-asympt}).

Near the critical line the interval $(-B,B)$ closes to a point and from
(\ref{integ0}) the dressed charge becomes 1.
Otherwise, for all values of $h$ we have solved the integral equation for the
dressed charge, Eq.~(\ref{integ0}), numerically. 
A plot of $\zeta(\theta)$, only inside the interval $(-B,B)$, is shown on
Fig.~\ref{dresscharge}.  The value of $\zeta(B)$ %to 1 as $B\rightarrow 0$,
decreases (increases) fast near $B=0$ for $\tau < 1/2$ ($\tau > 1/2$). For 
larger values of $B$, $\zeta(B)$  reaches quickly values close to its 
asymptotic value $\sqrt{2 \tau}$.

In the fermionic sector corresponding to the sine-Gordon model the 
chiral components $R_\up, L_\up$ of the fermion operator $\psi_\up$ are  given 
by the expressions 
\bea
R_\up & = &\exp\left[-\ri\sqrt{\pi/2}(\zeta + \zeta^{-1})\phi_\rc 
 - \right.\\ [1.5mm] \nonumber
&&\hspace{20mm}-\left.\ri\sqrt{\pi/2}(\zeta - \zeta^{-1})\bar\phi_\rc - 
\ri\sqrt{2\pi} \phi_\rs\right] 
\quad ,
\\ [3mm]
L_\up &= & \exp\left[\, \, \ri\sqrt{\pi/2}(\zeta - \zeta^{-1})\phi_\rc \right.+ 
\\[1.5mm] \nonumber
&&\hspace{20mm} + \left. \ri\sqrt{\pi/2}(\zeta + \zeta^{-1}) \bar\phi_\rc 
+  \ri\sqrt{2\pi} \bar\phi_\rs\right] 
\quad ,
\eea
where $\phi_\rc = (\Phi_\rc + \Theta_\rc)/2,\, \bar\phi_\rc = (\Phi_\rc - 
\Theta_\rc)/2$ and similarly for $\phi_\rs$.
The  correlation function of the right chiral component is given by 
\bea
\left<\left< R_\up(x,\tau) R^+_\up(0,0)\right>\right> = \frac{1}{
\left(v_\rc \tau - \ri x \right)^\frac{1}{2} } &&
\left(\frac{a^2}{v_\rc^2 \tau^2 + x^2}\right)^{\theta_\rc/2}\hspace{-4mm}\times
\nonumber \\ [2mm]
& & \hspace{-2mm} \times  \frac{1}{
\left(v_\rs \tau - \ri x \right)^\frac{1}{2} }
\, ,
\label{coreltn}
\eea

\begin{figure}
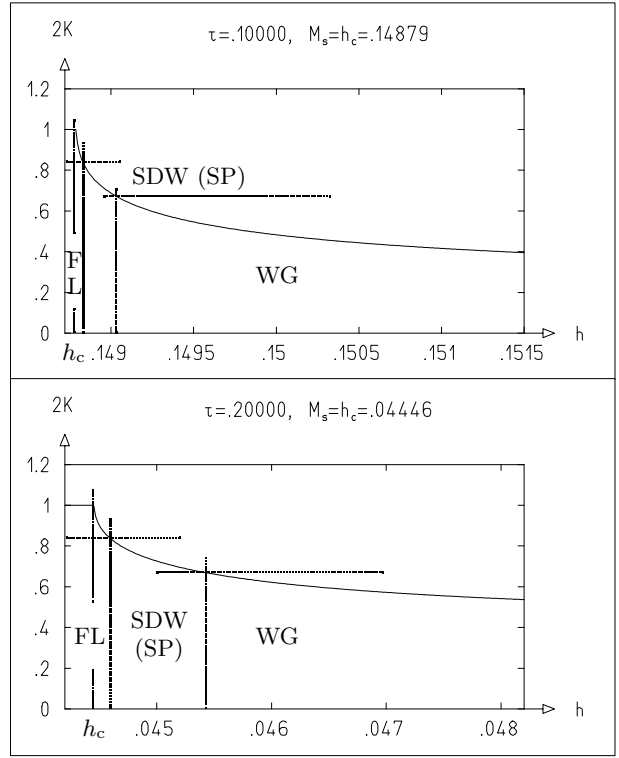

\unitlength=1mm
\begin{picture}(80,103)
%\put(-1.5,3){\epsfig{file=IQK2h.MET,height=50mm}}
\put(-1.5,53){\epsfig{file=IQ315.MET,height=50mm}}
\put(-1.5,3){\epsfig{file=IQ317.MET,height=50mm}}
\put(22.3,79.8){\makebox(0,0)[cc]{{\small{\small{\small{\small SDW (SP)}}}}}}
\put(7.1,69){\makebox(0,0)[cc]{{\small{\small{\small{\small F}}}}}}
\put(7.1,65.5){\makebox(0,0)[cc]{{\small{\small{\small{\small L}}}}}}
\put(34,67){\makebox(0,0)[cc]{{\small{\small{\small{\small WG}}}}}}
\put(11,77.4){\dashbox{0.2}(30,0)}
\put(6.05,82){\dashbox{0.2}(7,0)}
\put(12.6,59.3){\dashbox{0.2}(0,19)}
\put(8.2,59.3){\dashbox{0.2}(0,25.2)}
\put(7,72.5){\dashbox{0.2}(0,15)}
\put(7,59.4){\dashbox{0.2}(0,3)}
\put(7,56.6){\makebox(0,0)[cc]{{\small $h_\rc$}}}
\put(34,19){\makebox(0,0)[cc]{{\small{\small{\small{\small WG}}}}}}
\put(18.3,21){\makebox(0,0)[cc]{{\small{\small{\small{\small SDW}}}}}}
\put(18.3,17){\makebox(0,0)[cc]{{\small{\small{\small{\small (SP)}}}}}}
\put(9.1,19){\makebox(0,0)[cc]{{\small{\small{\small{\small FL}}}}}}
\put(6.05,32){\dashbox{0.2}(15,0)}
\put(18.1,27.4){\dashbox{0.2}(30,0)}
\put(11.8,9.4){\dashbox{0.2}(0,25)}
\put(24.5,9.3){\dashbox{0.2}(0,20)}
\put(9.5,23.5){\dashbox{0.2}(0,14.8)}
\put(9.5,9.4){\dashbox{0.2}(0,5)}
\put(9.7,6.6){\makebox(0,0)[cc]{{\small $h_\rc$}}}
%
%----- The following is the old version ------------
%\put(-1.5,53){\epsfig{file=IQ115.MET,height=50mm}} 115->315
%\put(-1.5,3){\epsfig{file=IQ117.MET,height=50mm}}  117->317
%\put(3.8,99.7){\makebox(0,0)[cc]{{\small {\small 2}}}}
%\put(3.8,49.7){\makebox(0,0)[cc]{{\small {\small 2}}}}
%%\put(-1.5,3){\epsfig{file=IQK-h.MET,height=50mm}}
%%\put(36,22){\epsfig{file=K-h.eps,height=18mm}}
%%\put(34.84,21.5){\epsfig{file=IQK2h.MET,height=20mm}}
%%\put(34,21){\line(0,1){21}}
%%\put(34,21){\line(1,0){33}}
%\put(6.05,83){\dashbox{0.2}(10,0)}
%\put(16,80.5){\makebox(0,0)[cc]{{\small{\small{\small{\small SDW}}}}}}
%\put(7.7,78){\dashbox{0.2}(30,0)}
%\put(26,73){\makebox(0,0)[cc]{{\small{\small{\small{\small WG}}}}}}
%\put(6.05,33){\dashbox{0.2}(10,0)}
%\put(16,30.5){\makebox(0,0)[cc]{{\small{\small{\small{\small SDW}}}}}}
%\put(8.1,28){\dashbox{0.2}(30,0)}
%\put(26,21){\makebox(0,0)[cc]{{\small{\small{\small{\small WG}}}}}}
%\put(6.9,59.3){\dashbox{0.2}(0,25.2)}
%\put(8.4,59.3){\dashbox{0.2}(0,20)}
%\put(6.5,9.4){\dashbox{0.2}(0,25)}
%\put(10.1,9.3){\dashbox{0.2}(0,20)}
%
\end{picture}
\caption{The dependence of the Luttinger parameter $K$ on the field
$h$ for $h>h_\rc$. The coupling constant $\tau$ is taken to have
values 0.1 and 0.2 respectively.
In the regions $K<1/3$, $1/3<K<\sqrt 2-1$ and $\sqrt 2-1<K<1/2$ the system is in
the Wigner crystal (WG), spin density
(SDW) insulating phases  and Fermi liquid phase respectively.}
\label{K-h}
\end{figure}

\hspace{-4mm}with \cite{Meden}
\be
\theta_\rc=\frac{1}{4}\left(K_\rc^{-1}+K_\rc-2\right)
\quad .
\label{theta}
\ee

In Eqs.~(\ref{coreltn}, \ref{theta}) one should not confuse $\tau$ and
the exponent $\theta$ with the coupling constant and the argument of the
integral equations used previously. 

From Fig.~\ref{dresscharge} and Eq.~(\ref{integ00}) it is clear that for a given
value of the coupling constant, the Luttinger parameter $K$ inside the
incommensurate phase changes its value
depending on the field $h$ (or on the number of solitons). Near the critical 
line as $h \rightarrow h_\rc^+(\tau)$, the Luttinger parameter approaches 1/2 
(for all the values of the coupling constant $\tau$ $(=\beta^2/8\pi)$, 
%the case $\beta \rightarrow 0$ is treated separately below) 
and from Eq.~(\ref{theta}) $\theta \rightarrow 1/8$, (here and in the
following we use $\theta$
instead of $\theta_\rc$). For large values of the
field $h$, $h \rightarrow \infty$, $K \rightarrow  \tau$ and therefore
$\theta \rightarrow (\tau-1)^2/4\tau$. Summarizing, for $K$ and $\theta$ we
have the following limits:
\begin{equation}
\left\{ \displaystyle \begin{array}{ll}\displaystyle
K \rightarrow 1/2 , \quad\quad \theta \rightarrow 1/8
\quad\quad\quad\mbox{($h \rightarrow h_\rc^+(\tau)$)}\ ;
\displaystyle\quad \\
\\
\displaystyle
K \rightarrow  \tau , \quad \theta \rightarrow (\tau-1)^2/4\tau
\quad\mbox{($h \rightarrow + \infty$)}\  .
\end{array}
\right.
\label{K-thet}
\end{equation}

For $\theta>1$ the single-fermion density of states (DOS)

%\begin{figure}
%\unitlength=1mm
%\begin{picture}(80,53)
%%\put(46,9){\makebox(0,0)[cc]{(a)}}
%%\put(-1.5,3){\epsfig{file=IQK+h.MET,height=50mm}}
%%\put(-1.5,3){\epsfig{file=IQ113.MET,height=50mm}}
%\put(-1.5,3){\epsfig{file=IQ117.MET,height=50mm}}
%\end{picture}
%\caption{The dependence of the Luttinger parameter $K$ on the field
%$h$ for $h>h_\rc$.}
%\end{figure}

\begin{figure}
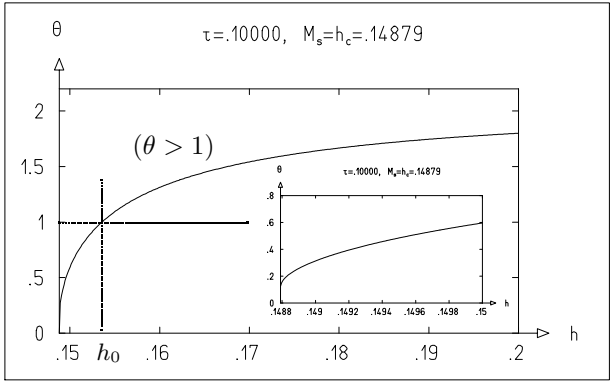

\unitlength=1mm
\begin{picture}(80,53)
%%\put(-1.5,3){\epsfig{file=IQ200.MET,height=50mm}}
%\put(-1.5,3){\epsfig{file=IQ202.MET,height=50mm}}
\put(-1.5,3){\epsfig{file=IQ302.MET,height=50mm}}
%%\put(5.9,25){\epsfig{file=IQ201.MET,height=17mm}}
%\put(32.8,10.5){\epsfig{file=IQ201.MET,height=20mm}}
%%\put(32.8,10.5){\epsfig{file=IQ301.MET,height=20mm}}
%\put(32.8,10.5){\epsfig{file=IQ501.MET,height=20mm}}
\put(5.8,23.9){\dashbox{0.2}(25,0)}
\put(11.4,9.6){\dashbox{0.2}(0,20)}
\put(12.4,6.6){\makebox(0,0)[cc]{{\small $h_0$}}}
\put(21,34){\makebox(0,0)[cc]{ $({\rm \theta} >1)$ }}
%\put(32.8,10.5){\epsfig{file=IQ401.MET,height=20mm}}
\put(32.,10.5){\epsfig{file=IQ601.MET.new,height=22mm}}
\end{picture}
\caption{Plot of the dependence of the exponent $\theta$ on the field
$h$ obtained for $h>h_\rc$. As it can be seen the
exponent $\theta$ which enters also in the single-particle density of
states DOS, increases monotonically with increasing $h$, and in case
$\tau<\tau_\rc=0.17157$ for values of $h$ bigger than a critical value
$h_0$ the exponent $\theta$
becomes larger than 1. In this region DOS exhibits pseudogap type
behavior.  In the inset is shown the square root dependence of $\theta$ on
$h$ in the region close to CL.}
\label{Thet-h}
\end{figure}

\hspace{-4mm}exhibits
pseudogap-type behavior. From Eq.~(\ref{K-thet}) this happens only when
$\tau < \tau_\rc $, where
\be
\tau_\rc= (3-2\sqrt{2}) \approx 0.17157
\quad .
\ee
For all values of $h$ in the interval $(h_\rc,+\infty)$ we find $\theta$
numerically.

Plots of the dependence of the Luttinger parameter $K$ and the
exponent $\theta$ on the field $h$ are represented on Fig.~\ref{K-h}
and Fig.~\ref{Thet-h} .

{\bf $(\beta \sim 1$, $\beta^2 < 8\pi)$ Behavior of $K$, $\theta$, 
$n_{\rm sol}$,
$\chi$ near the critical line (CL).} 
It is interesting to have the $h$ dependence of the above quantities
near the commensurate-incommensurate transition as $h\rightarrow h_\rc^+$. 
This can be achieved by solving the integral equation (\ref{integ}) by 
iterations. 
After two iterations the
boundary $B$ can be written as a series in powers of $(h/h_\rc-1)$
{\small
\bea
B= & \sqrt{2} & \left(\frac{h}{h_\rc} - 1\right)^{1/2}
\left[1-G(0)\frac{2\sqrt{2}}{3}\left(\frac{h}{h_\rc}-1\right)^{1/2} -
\right. \nonumber \\[3mm]
& - & \left. \left(\frac{1}{12}-\frac{20}{9}G(0)^2 \right)
\left(\frac{h}{h_\rc}-1\right) + \cdots
\right] \quad ,
\label{B-exp}
\eea
}
with coefficients which, up to the order shown, do not change in higher order 
iterations.
%This iterative process will only renormalize the coefficient of the
%least relevant power Each
In this way we get for the soliton contribution to the ground state energy:
{\small
\bea
\nonumber
{\cal E}_0  =  -  \frac{2\sqrt{2}M_\rs^2}{3\pi} 
& & \left(\frac{h}{h_\rc}-1\right)^{3/2} 
\left[1-2\sqrt{2}G(0)\left(\frac{h}{h_\rc}-1\right)^{1/2}  \right.
\\ [3mm] 
&& \left. 
+ \left(\frac{3}{20}+\frac{28 G^2(0)}{3}\right)\left(\frac{h}{h_\rc}-1\right)
\cdots
\right] \ ,
\label{expans1}
\eea
}
and for the dressed charge:

\begin{figure}
\unitlength=1mm
\begin{picture}(80,103)
%\begin{picture}(80,53)
%%\put(-3,3){\epsfig{file=IQNsH.MET,height=50mm}}
%%\put(-3,3){\epsfig{file=n_sol-h.MET,height=50mm}}
\put(-1.5,53){\epsfig{file=IQ210.MET,height=50mm}}
%\put(-1.5,53){\epsfig{file=n-sol-h2.MET,height=50mm}}
%
%\put(35,10.5){\epsfig{file=n_sol.eps,height=20mm}}
\put(35,60.5){\epsfig{file=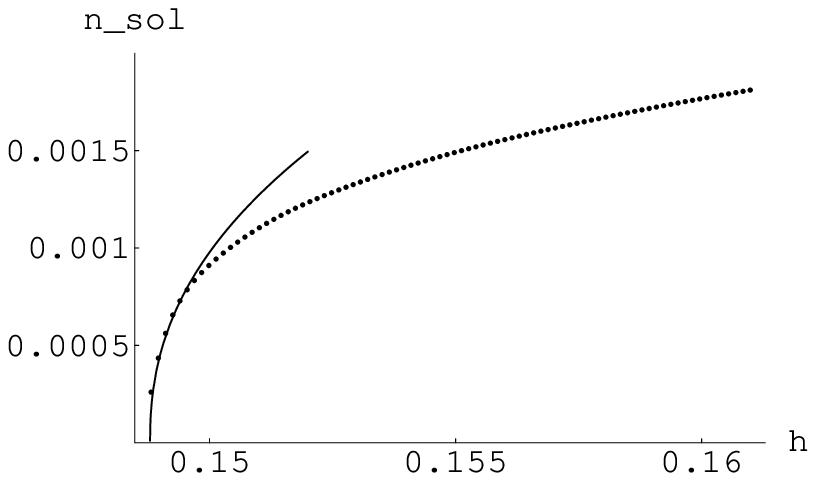,height=20mm}}
\put(-1.5,3){\epsfig{file=IQ205.MET,height=50mm}}
%\put(38,10.5){\epsfig{file=n_sol5.eps,height=18mm}}
\put(38,10.5){\epsfig{file=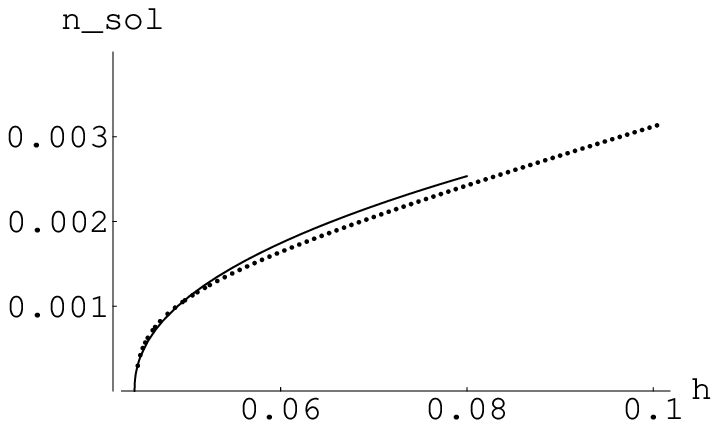,height=18mm}}
%%%\put(-3,3){\epsfig{file=IQnso.MET,height=50mm}}
%%%%\put(78,3){\epsfig{file=Btau2.eps,height=50mm}}
%%%\put(78,3){\epsfig{file=IQns2.MET,height=50mm}}
%\put(-1.5,53){\epsfig{file=IQnsh.MET,height=50mm}}
%\put(72,48){\makebox(0,0)[cc]{(b)}}
%\put(72,98){\makebox(0,0)[cc]{(a)}}
\end{picture}
\caption{Plots of the dependence of the number of
solitons on the magnetic field (chemical potential) $h$. The insets
compare the numerical (dotted line) and the analytic results near CL.
The coupling constant $\tau$ is taken to have values
$0.1$ and $0.2$, respectively.}
\label{nsol-h}
\end{figure}

\bea
\zeta(B) =  1 & - & 2\sqrt{2}G(0) \left(\frac{h}{h_\rc} - 1\right)^{1/2}
+
\nonumber \\ [3mm]
&+& \frac{32 G^2(0)}{3} \left(\frac{h}{h_\rc} -
1\right) - \cdots \quad .
\label{expans2}
\eea

Formulae (\ref{expans1}) and (\ref{chih}) show that (to first 
order) the
number of solitons and
the susceptibility have  square root dependence and square root
singularity as
functions of $h$, respectively.

Eq.~(\ref{expans2}) and Eq.~(\ref{integ00}) show, on the other hand,
that the Luttinger parameter $K$  decreases, from $K=1/2$ at $h=h_\rc$, as a 
square root function of $(h/h_\rc-1)$, whereas the exponent $\theta$ increases 
linearly having value 0 at $h=h_\rc$. Explicitly, $K$ and $\theta$ as 
functions of $(h/h_\rc-1)$, near the critical line, are given by:
\begin{equation}
\left\{ \displaystyle \begin{array}{ll}\displaystyle
K = \frac{1}{2} - 2\sqrt{2}G(0) \left(\frac{h}{h_\rc}-1\right)^{1/2} +
\cdots  \quad ,
%\quad\quad\quad\mbox{($h \rightarrow h_\rc^+(\tau)$)}\ ;
\displaystyle\quad
\\[3mm]
\displaystyle
\theta = \frac{1}{8} + \frac{3\sqrt{2}}{2} G(0)
\left(\frac{h}{h_\rc}-1\right)^{1/2} + \cdots
%\quad\mbox{($h \rightarrow \infty$)}\  .
\quad .
\end{array}
\right.
\label{K-thet-h}
\end{equation}
As functions of the number of solitons
\be
n_{\rm sol} \approx \frac{\sqrt{2}M_\rs}{\pi}
\left(\frac{h}{h_\rc}-1\right)^{1/2}+\cdots
\quad ,
\label{nsol}
\ee
they (to a first approximation) are given by:

%\vspace{1.2mm}
\begin{equation}
\left\{ \displaystyle \begin{array}{ll}\displaystyle
K = \frac{1}{2} - 2 \pi \frac{G(0)}{M_\rs}\ n_{\rm sol} \quad  
\quad\mbox{($n_{\rm sol} \rightarrow 0$)}\ ;
\displaystyle\quad 
\\[3.5mm]
\displaystyle
\theta = \frac{1}{8} +3 \pi \frac{G(0)}{2 M_\rs}\ n_{\rm sol}
\quad\quad\quad\mbox{($n_{\rm sol} \rightarrow 0$)}
\quad .
\end{array}
\right.
\label{K-thetn}
\end{equation}

%(h/h_\rc-1)$, whereas on the number of solitons the dependence is parabolic
%$\theta \sim 16 G(0)^2 \  n_{\rm sol}^2$.
A plot of the dependence of $\theta$ on the field $h$, 
near the critical line, is shown on Fig.~\ref{Thet-h}. The inset, in 
Fig.~\ref{Thet-h}, 
confirms the square root $h$-dependence of $\theta$ in that region.

%\begin{figure}
%\unitlength=1mm
%\begin{picture}(80,103)
%%\begin{picture}(80,53)
%%%\put(-3,3){\epsfig{file=IQNsH.MET,height=50mm}}
%%%\put(-3,3){\epsfig{file=n_sol-h.MET,height=50mm}}
%\put(-1.5,53){\epsfig{file=IQ210.MET,height=50mm}}
%%\put(-1.5,53){\epsfig{file=n-sol-h2.MET,height=50mm}}
%%
%%\put(35,10.5){\epsfig{file=n_sol.eps,height=20mm}}
%\put(35,60.5){\epsfig{file=n_sol2.eps,height=20mm}}
%\put(-1.5,3){\epsfig{file=IQ205.MET,height=50mm}}
%%\put(38,10.5){\epsfig{file=n_sol5.eps,height=18mm}}
%\put(38,10.5){\epsfig{file=n_sol6.eps,height=18mm}}
%%%%\put(-3,3){\epsfig{file=IQnso.MET,height=50mm}}
%%%%%\put(78,3){\epsfig{file=Btau2.eps,height=50mm}}
%%%\put(78,3){\epsfig{file=IQns2.MET,height=50mm}}
%\put(-1.5,53){\epsfig{file=IQnsh.MET,height=50mm}}
%\put(72,48){\makebox(0,0)[cc]{(b)}}
%%\put(72,98){\makebox(0,0)[cc]{(a)}}
%\end{picture}
%\caption{Plots of the dependence of the number of
%solitons on the magnetic field (chemical potential) $h$. The insets
%compare the numerical and the analytic results near CL. 
%The coupling constant $\tau$ is taken to have values
%$0.1$ and $0.2$, respectively.}
%\label{nsol-h}
%\end{figure}

\begin{figure}
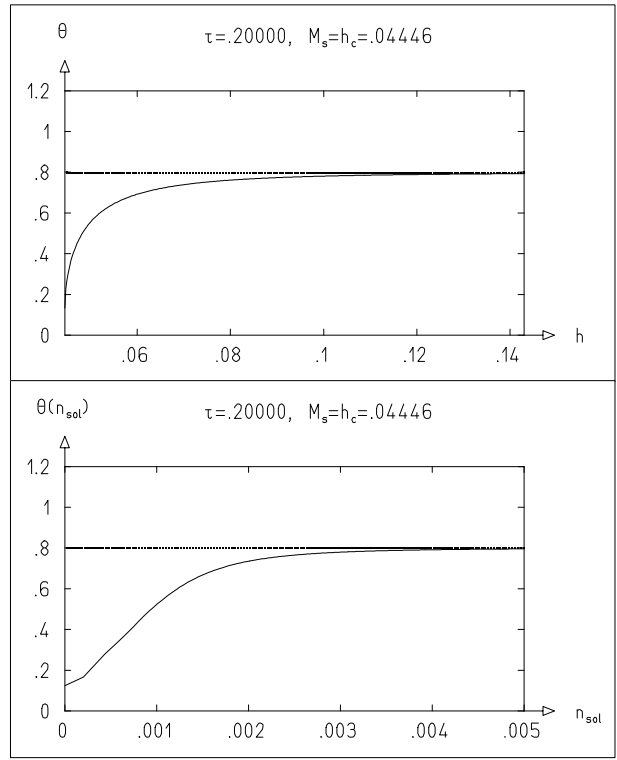

\unitlength=1mm
\begin{picture}(80,103)
%%\put(-1.5,3){\epsfig{file=IQTH1.MET,height=50mm}}
%% \put(-1.5,3){\epsfig{file=IQ211.MET,height=50mm}}
%%%\put(-3,3){\epsfig{file=IQchQ.MET,height=50mm}}
%%\put(-1.5,53){\epsfig{file=IQTH2.MET,height=50mm}}
%%\put(-1.5,53){\epsfig{file=IQ207.MET,height=50mm}}
%%\put(-1.5,53){\epsfig{file=new5.eps,height=60mm}}
%%\put(-2.3,51.5){\epsfig{file=new9.eps,height=53mm}}
%%\put(-2.07,50.2){\epsfig{file=new8.eps,height=53mm}}
%%\put(-2.07,50.2){\epsfig{file=new11.eps,height=53mm}}
%%\put(-2.07,50.2){\epsfig{file=new12.eps,height=53mm}}
%%\put(-2.07,50.2){\epsfig{file=new13.eps,height=53mm}}
%%\put(-2.07,50.2){\epsfig{file=new14.eps,height=53mm}}
%\put(-2.07,50){\epsfig{file=new15.eps,height=53mm}}
\put(-1.5,53){\epsfig{file=IQTheta-h.MET,height=50mm}}
%%\put(-1.5,3){\epsfig{file=IQ213.MET,height=50mm}}
%%\put(-1.5,3){\epsfig{file=IQ313.MET,height=50mm}}
\put(-1.5,3){\epsfig{file=IQ413.MET,height=50mm}}
%%%%%\put(78,3){\epsfig{file=IQch3.MET,height=50mm}}
%%%\put(6,49){\makebox(0,0)[cc]{(a)}}
%%%\put(87,49){\makebox(0,0)[cc]{(b)}}
\put(6,30.89){\dashbox{0.2}(60.8,0)}
\put(6,80.8){\dashbox{0.2}(60.8,0)}
\end{picture}
\caption{Plots of the dependence of the exponent
$\theta=1/8(\zeta(B)-2/\zeta(B))^2$, Eq.~(\ref{theta}), on the the field
$h$ and on the number of solitons respectively. For large $h$, the exponent
$\theta$ approaches its asymptotic value, $\theta\rightarrow (\tau-1)^2/4\tau$,
shown also in the plot. % given in Eq.~(\ref{K-thet}).
The parameter $\tau$ is taken to have value $0.2$ in both plots.}
%% DOS exhibits pseudogap-type behavior only for $\theta >1$,
%%i.e. above a certain (critical) number of solitons.}
\label{DOS}
\end{figure}

\begin{figure}
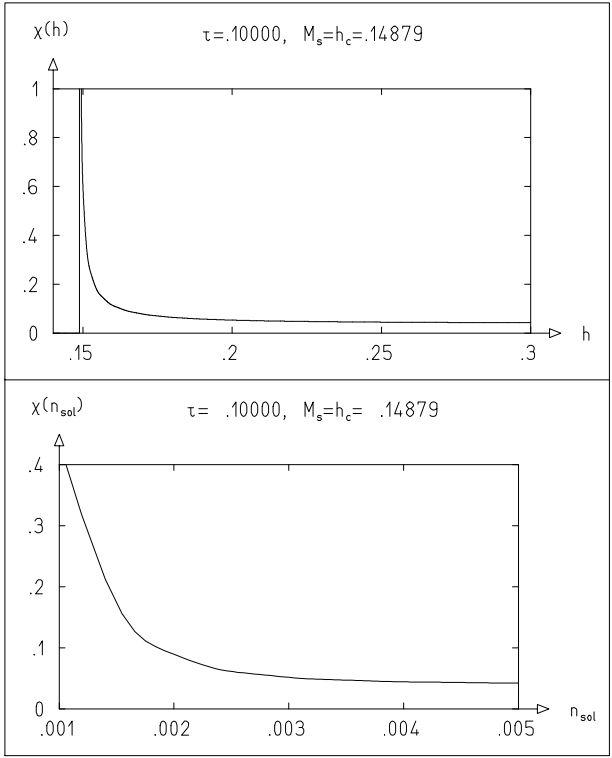

\unitlength=1mm
\begin{picture}(80,103)
%%%\put(-3,3){\epsfig{file=IQchq.MET,height=50mm}}
%%\put(-2,3){\epsfig{file=IQCHI.MET,height=50mm}}
%\put(-1.5,3){\epsfig{file=IQchN.MET,height=50mm}}
%\put(-1.5,3){\epsfig{file=IQChN.MET,height=50mm}}
%\put(-1.5,3){\epsfig{file=IQ218.MET,height=50mm}}
\put(-1.5,3){\epsfig{file=IQ219.MET,height=50mm}}
%\put(-1.5,3){\epsfig{file=IQhh3.MET,height=50mm}}
%%\put(78,3){\epsfig{file=IQch4.MET,height=50mm}}
%%\put(78,3){\epsfig{file=IQCH2.MET,height=50mm}}
%\put(-1.5,53){\epsfig{file=IQhh2.MET,height=50mm}}
%\put(-1.5,53){\epsfig{file=IQ214.MET,height=50mm}}
\put(-1.5,53){\epsfig{file=IQ215.MET,height=50mm}}
%\put(36,99){\makebox(0,0)[cc]{(\hspace{10mm})}}
\end{picture}
\caption{Plots of the dependence of the magnetic susceptibility $\chi$
on the magnetic field (chemical potential) $h$ (upper plot), and on
the number of solitons (lower plot), near the
critical line
are shown.  $\tau$ is taken to have value $0.1$ in both plots. In the
lower plot $\chi$ is shown only outside the region where it diverges
like
$1/n_{\rm sol}$ (see inset of Fig.~\ref{nsol-h} for $\tau=0.1$).}
\label{CHI-Nsol}
\end{figure}

In Fig.~\ref{nsol-h} we have represented the dependence of the number of
solitons, present in the incommensurate phase, on the field $h$ for
different values
of the coupling constant $\beta^2/8\pi$. Notice in (\ref{nsol}) that the
constant of proportionality of the number of solitons on $h$ grows for
decreasing $\tau$. This can be observed also numerically in the above
plots. In Fig.~\ref{DOS} we have represented the dependence of the exponent
$\theta$ on the number of solitons.
In Fig.~\ref{CHI-Nsol} we have represented the dependence of
susceptibility on the number of solitons, outside the region, near
the critical line, where it behaves like $\chi\sim 1/n_{\rm sol}$. Both
plots are obtained numerically, with $h$ used as parameter.

At small $\tau$, $G(0)$ behaves like $(1/\pi^2 \tau) \ln(1/\tau)$
and the expansions (\ref{expans1}) and (\ref{expans2}) are valid only for 
values of $h$ which fulfill
\be
\pi^2 \tau /\ln(1/\tau)\gg \left(h/h_\rc-1\right)
\quad .
\ee
%\begin{figure}
%\unitlength=1mm
%\begin{picture}(80,103)
%%\begin{picture}(80,53)
%%%\put(-3,3){\epsfig{file=IQNsH.MET,height=50mm}}
%%%\put(-3,3){\epsfig{file=n_sol-h.MET,height=50mm}}
%\put(-1.5,53){\epsfig{file=IQ210.MET,height=50mm}}
%%\put(-1.5,53){\epsfig{file=n-sol-h2.MET,height=50mm}}
%%
%%\put(35,10.5){\epsfig{file=n_sol.eps,height=20mm}}
%\put(35,60.5){\epsfig{file=n_sol2.eps,height=20mm}}
%\put(-1.5,3){\epsfig{file=IQ205.MET,height=50mm}}
%%\put(38,10.5){\epsfig{file=n_sol5.eps,height=18mm}}
%\put(38,10.5){\epsfig{file=n_sol6.eps,height=18mm}}
%%%%\put(-3,3){\epsfig{file=IQnso.MET,height=50mm}}
%%%%%\put(78,3){\epsfig{file=Btau2.eps,height=50mm}}
%%%%\put(78,3){\epsfig{file=IQns2.MET,height=50mm}}
%%\put(-1.5,53){\epsfig{file=IQnsh.MET,height=50mm}}
%%\put(72,48){\makebox(0,0)[cc]{(b)}}
%%\put(72,98){\makebox(0,0)[cc]{(a)}}
%\end{picture}
%\caption{Plots of the dependence of the number of
%solitons on the magnetic field (chemical potential) $h$. In (b) is
%represented the dependence of the number of solitons on $h$ near the
%critical line. The coupling constant $\tau$ is taken to have values 
%$0.100$ in both plots.}
%\label{nsol-h}
%\end{figure}

For smaller values of $\tau$ the expansion (\ref{expans1}) holds only in a
decreasingly small interval $h$ above $h_\rc$. For values of $h$ outside this
small interval a crossover to a different behavior takes place.
In this case one
has to take into account the presence of the breathers  
in the ground state \cite{JNW,Caux}. 

{\bf $(\beta \rightarrow 0)$ The limit of small $\tau$, $M_{\rm s}\tau $ = 
const.}
At $\beta^2<4\pi$ in the sine-Gordon spectrum,
in addition to solitons, breathers (kink-antikink bound states) will appear. 
The mass spectrum of the breathers is described by \cite{dashen}:
\be
M_n = 2 M_\rs \sin\left(n\frac{\pi \beta^2}{16\pi-2\beta^2}\right)
\label{breathers}
\quad ,
\ee
with $n$ -- integer, taking the following values: 
\be
n=1,\cdots ,\left[\frac{8\pi}{\beta^2}-1\right]
\quad .
\ee
When $\beta\rightarrow 0$ the bound-state energy vanishes and at small
$\beta^2$ one can expand the sine in (\ref{breathers}) and obtains
\be
M_n=n \pi M_\rs \tau 
\quad .
\ee
As $\beta\rightarrow 0$ the soliton mass grows to infinity while
$\beta^2 M_\rs$ converges to a finite value \cite{PT1}. The linearity in 
the mass spectrum
means that the breathers instead of being in 

\hspace{-4mm}the $n$-th bound-state
are as $n$ unbound particles. This is the classical limit, where the
sine-Gordon becomes a Klein-Gordon model.
In this limit an infinite number of breathers will appear in the
spectrum, filling the gap to the soliton mass $M_\rs$.
This problem is discussed in \cite{Caux}, and we use in the following
those results.

The boundary $B$, in the region $\tau \ll B \ll 1$, is found to satisfy
the following integral equation \cite{Caux}:
\bea
\int_{-B}^{B}\hspace{-4mm}\rd v \ln |\coth(\frac{v}{2})| & &
\left[\sinh^2 \hspace{-1mm} B \, \cosh^2 \hspace{-1mm}v  -
\cosh^2\hspace{-1mm}B \,  \sinh^2 \hspace{-1mm}v
\right]^{1/2}=
\nonumber \\ [3mm]
& & =\frac{2 \tau \pi^2}{ \mu}(h-h_\rc)
=2 \pi \frac{(h-h_\rc)}{h_\rc}
\quad ,
\eea
which for $B$ will give:
\bea
& & B^2  \ln(1/B)  \approx \pi \frac{(h - h_\rc)}{h_\rc}
\quad .
%\nonumber \\[3mm]
\eea
(In the above equations $\mu$ stands for the first breather's mass
$\mu=\pi\tau M_\rs=\pi \tau h_\rc$).
The contribution to the ground state energy of the kink condensate is:
\bea
{\cal E}_0  =   -
\frac{\mu^2}{4\pi\tau}\sinh^2 B \approx - \frac{\pi \mu}{2}
\frac{(h - h_\rc)}{\ln[h_\rc/(\pi(h - h_\rc))]} .
\eea

From this we obtain the equations for the number of solitons and the
susceptibility:
\be
\left\{ \displaystyle \begin{array}{ll}\displaystyle
n_{\rm sol} \approx \frac{\pi \mu}{2\ln\left[h_\rc/(\pi(h -
h_\rc))\right]}
\hspace{3.4mm}   \mbox{($\tau \ll B \ll 1$)} \ ;
\displaystyle\quad \\
\\
\displaystyle
\chi \approx  \frac{\pi \mu}{2 (h-h_\rc)\ln^2\left[h_\rc/(\pi(h
- h_\rc))\right]} \quad .
% \mbox{($\tau \ll B \ll 1 $)}
\end{array}
\right.
\label{small-tau}
\ee
If we express $h_\rc/(h-h_\rc)$ in terms of the number of solitons,
\bea
%& &  n \approx \frac{\pi \mu}{2\ln\left[h_\rc/(\pi(h - h_\rc))\right]}
%\quad ,
%\\[3mm]
& &  \frac{h_\rc}{h - h_\rc}   \approx
\pi      \exp\left(\frac{\pi \mu}{2n_{\rm sol}}\right)
\quad ,
\eea
for the susceptibility we get:
\bea
%\\ [3mm]
%\chi & \approx & \frac{\pi \mu}{2 (h-h\rc)\ln^2\left[h_\rc/(\pi(h
%- h_\rc))\right]}
%\nonumber \\ [3mm]
\chi & \approx &
\frac{\pi\tau}{2} \left(\frac{2n_{\rm sol}}{\mu}\right)^2
\exp\left(\frac{\pi \mu}{2n_{\rm sol}}\right)
\quad .
\eea
These results agree  with those obtained by 
\cite{Haldane,JNW}.
The criterion $\tau \ll B \ll 1$ means therefore:
\begin{eqnarray}
\tau \ll B \approx  2 \left( \frac{n_{\rm sol}}{\pi \mu} \right)^{1/2}
\exp\left[- \left(\frac{\pi \mu}{4n_{\rm sol}}\right)\right] \ll 1 \quad .
\label{cond2}
%\label{cond}
\eea

In order to calculate the Luttinger parameter in this region, we would
need to have in addition to $\chi$ the value of the Fermi velocity
$v_{\rF}$. We achive that through the following calculations.

% What we found with J-S was the following (see Eq.~(18)):
The excitation spectrum in this region is given by \cite{Caux}
\begin{eqnarray}
\epsilon(\theta) = - \mu\sqrt{\sinh(B - \theta)\sinh(B + \theta)}
\quad ,
\end{eqnarray}
whereas the equation for the momentum is:
%We have not solved the equation for the momentum which is
\bea
& &\int_{-B}^B\frac{\rd u \, p(u)}{\sinh(u - v)}  = \pi \mu\cosh v +
\\ [3mm]
& & \hspace{7mm} +\, 2\pi n_{\rm sol}\ln \left|
\coth\left[\left(v + B \right)/2 \right]  
\coth\left[\left(v - B\right)/2 \right] \right| %+ \pi \mu\cosh v \ ,
\quad .
\nonumber
\eea
After the substitutions
\[
x = \tanh v \quad , \quad  Y(x) = \cosh v \, p(v) \quad ,
\]
we get
\begin{eqnarray}
P\int_{-b}^b\frac{\rd y \, Y(y)}{x - y} = g(x) \quad ,
\end{eqnarray}
where $b = \tanh B$ and
\[
g(x) = \frac{\pi \mu}{1 - x^2} -
\frac{2\pi n_{\rm sol}}{\sqrt{1 - x^2}}
\ln\left[\frac{\sqrt{1 - b^2} + \sqrt{1 - x^2}}
{- \sqrt{1 - b^2} + \sqrt{1 - x^2}}\right] \ .
\]
%
%\begin{figure}
%\unitlength=1mm
%\begin{picture}(80,106)
%%%%\put(-3,3){\epsfig{file=IQchq.MET,height=50mm}}
%%%\put(-2,3){\epsfig{file=IQCHI.MET,height=50mm}}
%\put(-1.5,3){\epsfig{file=IQchN.MET,height=50mm}}
%%\put(-1.5,3){\epsfig{file=IQhh3.MET,height=50mm}}
%%%\put(78,3){\epsfig{file=IQch4.MET,height=50mm}}
%%%\put(78,3){\epsfig{file=IQCH2.MET,height=50mm}}
%\put(-1.5,53){\epsfig{file=IQhh2.MET,height=50mm}}
%\end{picture}
%\caption{Plots of the dependence of the magnetic susceptibility $\chi$
%on the magnetic field (chemical potential) $h$ near the critical line
%are shown.  $\tau$ is taken to have values $0.100$ and $0.060$, respectively.}
%\label{CHI-Nsol}
%\end{figure}
%
The general solution of such equation is
\begin{eqnarray}
Y(x) = \frac{1}{\pi^2} \sqrt{b^2 - x^2} P\int_{-b}^b
\frac{g(y)\, \rd y}{(y - x)\sqrt{b^2 - y^2}} \quad .
\end{eqnarray}
The solution at small $b \approx B \ll 1$ is
\begin{eqnarray}
p_\rF - p = \frac{4 n_{\rm sol}}{b}\sqrt{b^2 - x^2}, ~~ \epsilon(x) = 
- \mu\sqrt{b^2 - x^2}\ ,
\end{eqnarray}
which gives the following dispersion law:
\bea
E = v_\rF(p - p_\rF)\quad ,
\eea
with
\bea
 \ v_\rF = \frac{b \mu}{4n_{\rm sol}} \approx 
\frac{1}{2}\left(\frac{\mu}
{\pi n_{\rm sol}}\right)^{1/2} 
\exp{\left(-\frac{\pi \mu}{4n_{\rm sol}} \right)} \quad .
%\re^{-\pi \mu/4n_{\rm sol}} \ .
\eea
The maximal value of $-\epsilon$ is $\mu\sinh B \approx \mu b$.
This means that $E(p)$ deviates from the linear form at $p \sim p_\rF$
indicating  that the ultraviolet  cut-off in the momentum space is of order of
$p_\rF \sim n_{\rm sol}$.

According to the above results in the region (\ref{cond2}) we have
\bea
K = \frac{1}{2} \pi\chi v_\rF&\approx& \frac{\tau}{2} 
\left(\frac{\pi n_{\rm sol}}{ \mu }\right)^{3/2}
\exp\left(\frac{\pi \mu}{4n_{\rm sol}}\right)
\nonumber \\ [3mm]
%&\ll&\sqrt{\pi} B (n_{\rm sol}/\pi \mu)^{3/2}\exp[(\pi \mu/4n_{\rm sol})]
%\nonumber \\ [3mm]
& \sim & \frac{1}{2} \left( \frac{n_{\rm sol}}{\pi \mu } \right)^{2}
\quad .
\label{K-nsol}
\eea
When $n_{\rm sol}$ approaches the boundary of validity of this expression 
given by 
Eq.~(\ref{cond2}), which corresponds to very small $n_{\rm sol}/\mu$, 
$K$ is still small: 
$ K \sim (n_{\rm sol}/\pi \mu)^{2}/2$. At even smaller $n_{\rm sol}$ it  
asymptotically
approaches 1/2. In this region Eq.~(\ref{zeta}) can be solved by iterations.

Thus at $\tau \ll 1$ (the classical regime)  the Luttinger parameter $K$ 
decreases quickly from 1/2 towards
its asymptotic value $\tau$. This rapid change takes place for $n_{\rm
sol} < \mu$, 
that is, it is determined not by the soliton mass which goes to infinity at 
$\beta^2 \rightarrow 0$, but by the mass of the breather which remains 
constant in this limit. 

%\newpage
%\vspace{5mm}

\begin{center}
\hspace{0mm}{\small {\bf DISCUSSION}  } 
\end{center}
%\vspace{3mm}

In this paper we have studied the scaling exponents of the isospin
sector of the U(1) Thirring model in the incommensurate phase when its
energy spectrum is gapless. This model may be used as an effective
theory for a low-energy limit of fermionic theory with long range
repulsion on a lattice close to half-filling.  Exactly at half-filling
 the charge
sector is gapped and is described by the sine-Gordon model with a relevant
$cos$ term. This gap is suppressed by a finite chemical potential and
the system goes into a gapless incommensurate phase. 
In this  phase, which is the phase of concern in
this paper, the correlation functions of the operators 
(\ref{staggered}-\ref{triplet}) and (\ref{coreltn}) decay
as power laws of some critical exponents which are given in terms of a
single parameter $K$, the Luttinger liquid parameter. $K$ is
related to
the dressed charge which satisfies an integral equation characteristic
for
exactly solvable models.

We are mainly interested in the region of small $\beta$. This is an
interesting limit which has not been much  discussed in literature.
At $\beta \rightarrow 0$  the solitons become infinitely heavy and in the
insulating phase (absence of solitons) SG model becomes a model of free
massive bosonic field with the mass $\mu$. It turns  out that even in
the incommensurate phase  when the chemical potential forces solitons to
appear,  $K$ 
does not depend on the soliton mass but on the mass of the optical phonon
$\mu$ (see Eq.~(\ref{K-nsol})).

 As we have shown, the value of $K$ depends on doping (that is the
soliton density) with  $K = 1/2$ in the limit of zero doping and
reaching its asymptotic value $\beta^2/8\pi$ at large
doping. Therefore changing the number of solitons one can explore different
phases of the model (see Fig. 3). Thus for small doping the
single-particle hopping is relevant. In this case one might expect that  the system of many
coupled chains will become  a Fermi liquid.  When doping increases $K$
decreases, the single particle tunneling processes are suppressed, but
the spin susceptibility becomes singular. This is a region when
interchain coupling will establish Spin Density Wave (SDW) ordering. For $K<1/3$ the $4 k_\rF$-component of the
charge density operator has the smallest dimension with the
corresponding susceptibility being the most singular. 
 This notifies that the Wigner
crystal order parameter will be  stabilized by a Coulomb interaction  in
a three dimensional array of coupled chains. 
 
The number of solitons and the susceptibility are found analytically and
numerically and  display, at the transition point, square-root behavior and 
square-root singularity respectively. 
As $\beta \rightarrow 0$, the above dependence holds only in a decreasingly
small interval $h$ above $h_\rc$. For larger $h$ the dependence of
$n_{\rm sol}$ and $\chi$ on $h$ is given in Eq.~(\ref{small-tau}). 
Plots of the dependence of $n_{\rm sol}$ 
and $\chi$ on $h$ are given on Fig.~\ref{nsol-h} and Fig.~\ref{CHI-Nsol}.

We are grateful to Alan Luther who encouraged us to study this problem. 
A.~M.~T. acknowledges the hospitality of Isaac Newton Institute where part
of this work was completed. E.~P. acknowledges discussions with Tilo Stroh.

\begin{center}
{\small {\bf APPENDIX}}
\end{center}

%\appendix{\underline{\underline{ {\bf APPENDIX }}}}
%\vspace{5mm}
%{\underline{\underline{ {\bf A 1:}}}} {\bf The Wiener-Hopf method}
%{\underline{\underline{ {\bf A 1:}}}} \hspace{2mm} 

\begin{center}
{\bf Expansion near the critical line}
\end{center}

The kernel $G(\theta)$ of (\ref{integ}) can be expanded about $\theta=0$
in a Taylor series with convergence radius $r_\theta=\pi\tau/(1-\tau)$.
For small $h_{\rm c}-h=M_{\rm s}-h$, using the Taylor expansion of the
inhomogeneity $h_{\rm c}\cosh(\theta)-h$, the integral equation can be
solved for a given accuracy by a power series about $\theta=0$ implying
also a series in $B$ about $B=0$:
\bgroup
\arraycolsep=0pt
\begin{eqnarray}
&&
G(\theta) \sim \sum_{i=0}^{N}\frac{G_{2i}}{(2i)!}\theta^{2i} \quad,\quad
G_{2i}=G^{(2i)}(0) \quad, \\
&&
h_{\rm c}\cosh(\theta)-h \sim h_{\rm c}\sum_{i=0}^{N}\frac{1}{(2i)!}
\theta^{2i}-h \quad,\quad \\
&&
\epsilon(\theta) \sim \sum_{i=0}^{N}\epsilon_{2i}\theta^{2i} \quad,\quad
\epsilon_{2i} \sim \sum_{j=0}^{2(N-i+1)}\epsilon_{2i,j}B^j \quad.
\quad\:\,
\label{eq:eps-expans}
\end{eqnarray}
\egroup

%With symbolic algebra programs used throughout the coefficients were
%determined. 
Beginning with $\epsilon_{2i,j}=0$ the iteration for the
coefficients converges after $2N+1$ cycles, where actually $N=4$ was
used and results for $N=2$ are shown below.
The solution for $B$ from the condition $\epsilon(B)=0$, where
$\epsilon(B)$ is now a sole power series in $B$, is parameterized as a
power series in $\sqrt{h/h_{\rm c}-1}$, cf.\ (\ref{B-exp}),
\begin{eqnarray}
B& \sim &\sum_{j=1}^{2N+3}\alpha_j x^j \quad,\quad
  x=\sqrt{\frac{h}{h_{\rm c}}-1} \quad,\quad
\label{eq:B-ser-x} \\*
\epsilon(B)& \sim &\sum_{i=0}^{2N+2}\epsilon^{(B)}_i B^i \sim
  \sum_{i=0}^{2N+4}\epsilon^{(h)}_i x^i \quad,
\label{eq:eps-B-ser}
\end{eqnarray}
where also $h$ is replaced by $h_{\rm c}(1+x^2)$ in $\epsilon^{(B)}_i$,
which yields $\epsilon^{(h)}_1=\epsilon^{(h)}_0=0$ and
$\epsilon^{(h)}_2=
h_{\rm c}(\alpha_1^2-2)/2$. From $\epsilon^{(h)}_{i+1}=0$,
$i=2,\ldots,2N+3$, the coefficients $\alpha_i$ can be read
off recursively.

In the soliton contribution to the ground-state energy (\ref{ground-st}) the
power expansion of $\cosh \theta$ and (\ref{eq:eps-expans}) is used, where the 
integration leads to a power series in $B$ and replacement (\ref{eq:B-ser-x}) 
to a series
in $\sqrt{h/h_{\rm c}-1}$:
\begin{equation}
{\cal E}_0 \sim -\frac{2\sqrt{2}M_{\rm s}^2}{3\pi}x^3\sum_{i=0}^{2N+2}
\tilde{\cal E}_0^{(i)}x^i \quad, \quad x=\sqrt{\frac{h}{h_{\rm c}}-1}
\quad.
\label{eq:E_0-exp-x}
\end{equation}

The integral equation (\ref{integ0}) for
$\zeta$ is treated exactly as (\ref{integ}), leading to a series in
$\theta$ and $B$. For $\zeta(B)$ applying the previous replacement for
$B$ the series reads
\begin{equation}
\zeta(B) \sim \sum_{i=0}^{2N+2}\zeta^{(h)}_i x^i \quad,\quad
x=\sqrt{\frac{h}{h_{\rm c}}-1}\quad.
\label{eq:zeta-exp-x}
\end{equation}
Finally, this series is replaced into (\ref{integ00}), which then is
inserted into (\ref{theta}), leading to the following series:
\begin{equation}
  K  \sim  \sum_{i=0}^{2N+2}K_i x^i \quad, \quad
  \theta  \sim  \sum_{i=0}^{2N+2}\theta_i x^i \quad.
\label{eq:K,theta-exp-x}
\end{equation}

\subsection*{Coefficients $\alpha_i$ of $B$ in (\ref{eq:B-ser-x})}
%\par\hangindent\parindent\hangafter=1\noindent
\begin{description}
\item
 \rightskip 0pt plus \hsize
 $\alpha_1  =  \sqrt{2}$,
\item
 \rightskip 0pt plus \hsize
 $\alpha_2  =  -4G_0/3$,
\item
 \rightskip 0pt plus \hsize
 $\alpha_3  =  \sqrt{2}(-1/12 + 20 G_0^2/9)$,
\item
 \rightskip 0pt plus \hsize
 $\alpha_4  =  2G_0/5 - 256 G_0^3/27 - 8 G_2/5$,
\item
 \rightskip 0pt plus 0.5\hsize
 $\alpha_5  =  \sqrt{2}(3/160 - 49 G_0^2/45 + 616 G_0^4/27 +
               256 G_0 G_2/45)$,
\item
 \rightskip 0pt plus \hsize
 $\alpha_6  =  8(-2 G_0/35 + 64 G_0^3/27 - 3584 G_0^5/81 +
               (29/35 - 128 G_0^2/3)G_2/3 -
               4 G_4/21)/3$
\item
 \rightskip 0pt plus \hsize
 $\alpha_7  =  \sqrt{2}(-5/896 + 769 G_0^2/1400 - 286 G_0^4/15 +
                77792 G_0^6/243  + 32(-193/35 +
               176 G_0^2) G_0G_2/45 + 272 G_2^2/75 +
               736 G_0 G_4/315)$
\end{description}

\subsection*{Coefficients $\tilde{\cal E}_0^{(i)}$ of ${\cal E}_0$
in (\ref{eq:E_0-exp-x})}
\begin{description}
\item
 \rightskip 0pt plus \hsize
 $\tilde{\cal E}_0^{(0)} = 1$
\item
 \rightskip 0pt plus \hsize
 $\tilde{\cal E}_0^{(1)} = -2\sqrt{2}G_0$
\item
 \rightskip 0pt plus \hsize
 $\tilde{\cal E}_0^{(2)} = 3/20 + 28 G_0^2/3$
\item
 \rightskip 0pt plus \hsize
 $\tilde{\cal E}_0^{(3)} = -2\sqrt{2}(G_0/5 + 320 G_0^3/27 + 8 G_2/15)$
\item
 \rightskip 0pt plus \hsize
 $\tilde{\cal E}_0^{(4)} = -3/224 + 9 G_0^2/5 + 1144 G_0^4/9 +
                     64 G_0 G_2/5$
\item
 \rightskip 0pt plus \hsize
 $\tilde{\cal E}_0^{(5)} = 2\sqrt{2}(G_0/28 - 16 G_0^3/9 - 14336
G_0^5/81 +
                     4(-1/35 - 64 G_0^2/9) G_2 - 4 G_4/35)$
\item
 \rightskip 0pt plus \hsize
 $\tilde{\cal E}_0^{(6)} = 1/384 - 2321 G_0^2/4200 + 286 G_0^4/27 +
                     1478048 G_0^6/729 + 16(61/175 +
                     2288 G_0^2/27) G_0 G_2/3 + 1232 G_2^2/225 +
                     352 G_0 G_4/105$
\end{description}

\subsection*{Coefficients $\zeta_i^{(h)}$ of $\zeta(B)$
in (\ref{eq:zeta-exp-x})}
\begin{description}
\item
 \rightskip 0pt plus \hsize
 $\zeta_0^{(h)}=1$
\item
 \rightskip 0pt plus \hsize
 $\zeta_1^{(h)}=-2\sqrt{2}G_0$
\item
 \rightskip 0pt plus \hsize
 $\zeta_2^{(h)}=32 G_0^2/3$
\item
 \rightskip 0pt plus \hsize
 $\zeta_3^{(h)}=\sqrt{2}(G_0/2 - 280 G_0^3/3 - 8 G_2)/3$
\item
 \rightskip 0pt plus \hsize
 $\zeta_4^{(h)}=32(-G_0^2/5 + 160 G_0^4/9 + 14 G_0 G_2/5)/3$
\item
 \rightskip 0pt plus \hsize
 $\zeta_5^{(h)}=\sqrt{2}(-3 G_0/80 + 154 G_0^3/15 - 16016 G_0^5/27 +
                2(1/3 - 336 G_0^2/5) G_2 - 16 G_4/15)$
\item
 \rightskip 0pt plus \hsize
 $\zeta_6^{(h)}=16(13 G_0^2/35 - 448 G_0^4/9 + 57344 G_0^6/27 +
                8(-29 G_0/35 + 704 G_0^3/9) G_2 + 58 G_2^2/5 +
                272 G_0 G_4/35)/9$
\end{description}

\subsection*{Coefficients $K_i$ of $K$ in (\ref{eq:K,theta-exp-x})}
\begin{description}
\item
 \rightskip 0pt plus \hsize
 $K_0=1/2$
\item
 \rightskip 0pt plus \hsize
 $K_1=-2\sqrt{2}G_0$
\item
 \rightskip 0pt plus \hsize
 $K_2=44 G_0^2/3$
\item
 \rightskip 0pt plus \hsize
 $K_3=(G_0 - 944 G_0^3/3 - 16 G_2)/3\sqrt{2}$
\item
 \rightskip 0pt plus \hsize
 $K_4=2(-7 G_0^2/5 + 5008 G_0^4/27 + 304 G_0 G_2/15)$
\item
 \rightskip 0pt plus \hsize
 $K_5=-3 G_0/80 + 734 G_0^3/45 - 35216 G_0^5/27 +
      2(1 - 5008 G_0^2/15) G_2/3 - 16 G_4/15$
\item
 \rightskip 0pt plus \hsize
 $K_6=8(11 G_0^2/35 - 8236 G_0^4/135 + 277600 G_0^6/81 +
      4(-151/7 + 27016 G_0^2/9)G_0 G_2/15 + 52 G_2^2/5 +
      712 G_0 G_4/105)/3$
\end{description}

% This is the case of fermions with spin

%The changes are: (a) The '\theta' sum now extends to "2N+2" instead to
%"2N+3". (b) All the coefficients changed.

\subsection*{Coefficients $\theta_i$ of $\theta$ in
(\ref{eq:K,theta-exp-x})}
\begin{description}
\item
 \rightskip 0pt plus \hsize
 $\theta_0=1/8$
\item
 \rightskip 0pt plus \hsize
 $\theta_1=3\sqrt{2}G_0/2$
\item
 \rightskip 0pt plus \hsize
 $\theta_2=5G_0^2$
\item
 \rightskip 0pt plus \hsize
 $\theta_3=\sqrt{2}(-G_0 + 16(- 7G_0^3 + G_2))/8$
\item
 \rightskip 0pt plus \hsize
 $\theta_4=(-51 G_0 + 16(535 G_0^3 + 69 G_2))G_0/90$
\item
 \rightskip 0pt plus \hsize
 $\theta_5=\sqrt{2}(81 G_0 + 32(355 G_0^3 - 9(5 G_2 - 8 G_4) -
      40(743 G_0^3 + 130 G_2)G_0^2))/2880$
\item
 \rightskip 0pt plus \hsize
 $\theta_6=2 (39 G_0^2 + 84(43 G_2^2 - 227 G_0^4) +
       32(33985 G_0^3+ 8001 G_2)G_0^3  -
       12(107 G_2 - 138 G_4)G_0)/945$
\end{description}

% The coefficients $\theta_i$ of $\theta$ for the case of fermions without
%                                                            spin

\subsection*{Coefficients $\theta_i$ of $\theta$ in
(\ref{eq:K,theta-exp-x}), for fermions without spin}
\begin{description}
\item
 \rightskip 0pt plus \hsize
 $\theta_0=\theta_1=0$
\item
 \rightskip 0pt plus \hsize
 $\theta_2=16 G_0^2$
\item
 \rightskip 0pt plus \hsize
 $\theta_3=-160\sqrt{2}G_0^3/3$
\item
 \rightskip 0pt plus \hsize
 $\theta_4=8(-G_0^2 + 140 G_0^4 + 16 G_0 G_2)/3$
\item
 \rightskip 0pt plus \hsize
 $\theta_5=8\sqrt{2}(91 G_0^3 - 7360 G_0^5 - 1264 G_0^2 G_2)/45$
\item
 \rightskip 0pt plus \hsize
 $\theta_6=32(G_0^2/5 - 137 G_0^4/3 + 23176 G_0^6/9 +
           4(-1 + 452 G_0^2/3) G_0 G_2 + 8 G_2^2 + 24 G_0 G_4/5)/9$
\item
 \rightskip 0pt plus \hsize
 $\theta_7=2\sqrt{2}(-613 G_0^3/70 + 10048 G_0^5/9 - 431744 G_0^7/9 +
           16(1103/35 - 8032 G_0^2/3) G_0^2G_2/3 -
           1408 G_0 G_2^2/3 - 5504 G_0^2 G_4/35)/3$
\end{description}

 Our results can be generalized for the case of spinless
fermions. Here  instead of Eq.~(\ref{K-chi}) and Eq.~(\ref{theta})
one has to use 
\be
K=\pi \chi v_\rF \quad, \quad \quad   \theta=\frac{1}{2}\left(\sqrt{K} -
\frac{1}{\sqrt K}\right)^2
\quad .
\ee

\vspace{5mm}

\end{document}